\documentclass[draft]{cimento}
\title{On 
 the vorticity of the Universe}
\author{D.~PALLE}
\instlist{\inst{}Zavod za teorijsku fiziku \\
Institut Rugjer Bo\v skovi\'c \\
Po\v st. Pret. 180, HR-10002 Zagreb, CROATIA}
\PACSes{\PACSit{98.80}{Cosmology}}
\begin{document}

\maketitle

\begin{abstract}
Recent analyses of the first-year WMAP data claim 
large-scale asymmetry and anisotropy of the CMBR
fluctuations.
We argue that the covariant and gauge invariant treatment
of density fluctuations formulated by Ellis and Bruni 
can explain the asymmetric and anisotropic WMAP data
by including the vorticity of the Universe.
It appears that the spatial gradients of the density contrast 
are proportional to the vorticity of the Universe,
thus allowing measurements and quantifications of the magnitude
and axis of the possible cosmological rotation.
\end{abstract}

The first-year WMAP data represent the first step
towards higher-precision cosmology.
As expected, the WMAP data confirmed some standard cosmological
settings concerning baryonic and nonbaryonic matter,
primordial fluctuations, cold dark matter paradigm or
spatial flatness.
We should also stress unexpected findings, such as
low large-scale power of fluctuations that cannot be fitted by 
flat Friedmann cosmologies with a positive cosmological constant
inferred from small-scale structure data, very small bound on tensor-mode
fluctuations (contrary to the expectations based on the
inflationary models), running of the power index at small
scales and rather large reionization redshift and optical depth \cite{r1}.
It is interesting to note that the integrated Sachs-Wolfe effect 
of the models
with the negative cosmological constant reduces the large scale power
of fluctuations.

Eventually, we should be forced to change both the physical 
content \cite{r2} and the geometry of the Universe \cite{r3} to fully understand
a detailed picture growing from more refined cosmic data.

The small amount of acceleration in the geometry beyond that
of Robertson-Walker can substantially change the evolution of
the density contrast at small redshifts \cite{r4}. 
For the same initial contrast at high redshift, 
Eq.(13) of Ref.[4] gives the same deviation
from $\Omega_{m}=1$ at zero redshift for the following cosmic parameters:
$\delta (\Omega_{m}=1,\Sigma=0)-\delta (\Omega_{m}=2, \Sigma=0) \simeq
\delta (\Omega_{m}=1, \Sigma=0)-\delta (\Omega_{m}=1, \Sigma=3\times 10^{-3})$,
where $\Omega_{m}+\Omega_{\Lambda}=1$ is assumed and $\Sigma$ is the
acceleration parameter $A_{c}A^{c}=\Sigma H^{2},\ A_{c}=\dot{u}_{c}$ \cite{r4}.
From such calculations one can conclude on the importance of studying 
more general geometries with the inclusion of acceleration and
vorticity of the Universe.
It is intriguing that cosmic acceleration induces anomalous
constant acceleration towards the Sun \cite{r5},that is much smaller than the 
Pioneer anomaly $a_{\Sigma}=\sqrt{\Sigma}cH_{0} << a(Pioneer)$.
However, the Pioneer anomaly is too large to be consistent with planetary 
orbits.

Heavy Majorana neutrinos emerging from nonanomalous and nonsingular
theory \cite{r2} are good candidates for cold-dark-matter particles
to solve the problem of the cosmic photon diffuse background \cite{r6},
large optical depth observed by WMAP or super GZK cut-off
cosmic rays \cite{r7}.
There is also the possibility that the galactic center point source
emission with no variability measured by Whipple \cite{r8} and Cangaroo II
\cite{r9} could be interpreted in terms of the CDM galactic annihilation
of very heavy particles with large annihilation cross sections \cite{r10}.
Future measurements of Cangaroo III, Magic, H.E.S.S. and Veritas
atmospheric \v Cerenkov telescopes will resolve this possible
indirect evidence for the CDM.

Recent analyses of non-Gaussianity  of WMAP data by wavelet techniques 
converge to the conclusion on positive non-Gaussian signals at
large scales. The first findings of Vielva et al \cite{r11}
were confirmed by Mukherjee and Wang \cite{r12}.
Using the local curvature method Cabella et al \cite{r13} were able
to reproduce observed non-Gaussian features.
The N-point correlation functions from the WMAP data \cite{r14}
also pointed to asymmetry at large scales.
It was also shown that the fit of cosmological parameters, such as 
optical depth or
the amplitude of fluctuations at two hemispheres, gave completely 
different values \cite{r15}.
The most convincing analysis was performed using
directional continuous spherical wavelets resulting in 
the $6\sigma$ detection of non-Gaussianity \cite{r16}.

It is very difficult to find any physical source of
such asymmetric WMAP data.
We try to understand it assuming the cosmic geometry with
vorticity and acceleration of the Universe.
Let the metric have the form of Obukhov and Korotky \cite{r17}:

\begin{eqnarray}
ds^{2}&=&-dt^{2}+R^{2}(t)(dx^{2}+(1-\Sigma)a^{2}(x)dy^{2}) \\
      &+&R^{2}(t)dz^{2}+2R(t)b(x)dydt, \nonumber
\end{eqnarray}
\begin{eqnarray*}
a(x)=e^{mx},\ b(x)=\sqrt{\Sigma}a(x),
\end{eqnarray*}
\begin{eqnarray*}
A_{b} \equiv u^{c}\nabla_{c}u_{b}=b\dot{R}\delta^{2}_{b},
\ \omega_{ab}=Rb'\delta^{1}_{[a}\delta^{2}_{b]},\ 
[ab]\equiv \frac{1}{2}(ab-ba).
\end{eqnarray*}

Ellis and Bruni formulated \cite{r18} a covariant and gauge-invariant
formalism for density fluctuations in inhomogeneous and nonisotropic
spacetimes. Careful reanalysis \cite{r19} found an extra term with
the spacelike gradient of vorticity in the evolution
equation for the density contrast (Eq.(26) of Ref. [19]):

\begin{eqnarray}
\ddot{{\cal D}}_{\perp a}+{\cal A}(t)\dot{{\cal D}}_{\perp a}
-{\cal B}(t){\cal D}_{a}+{\cal L}{\cal D}_{a}
-{\cal C}(t)^{(3)}\nabla^{b}\omega_{ab}=0,
\end{eqnarray}
\begin{eqnarray*}
{\cal D}_{a}\equiv R\frac{^{(3)}\nabla_{a} \rho}{\rho},\ 
^{(3)}\nabla_{a} f=h_{a}^{b}\nabla_{b} f, \\
h_{ab}=g_{ab}+u_{a}u_{b},\ u^{a}u_{a}=-1, \\
for\ further\ definitions\ see\ Ref.[19].
\end{eqnarray*}

This additional term vanishes for the 
cosmic model of Obukhov and Korotky (Eq.(1)), thus
we have to try another approach.
It is possible to make the following local decomposition
of the spatial variation of the density contrast \cite{r19}:

\begin{eqnarray}
R^{(3)}\nabla_{b}{\cal D}_{a}\equiv \Delta_{ab}=W_{ab}+\Sigma_{ab}
+\frac{1}{3}\Delta h_{ab},
\end{eqnarray}
\begin{eqnarray*}
W_{ab}\equiv \Delta_{[ab]},\ \Sigma_{ab}\equiv
\Delta_{(ab)}-\frac{1}{3}\Delta h_{ab},\\
\Sigma_{ab}=\Sigma_{(ab)},\ \Sigma^{a}_{a}=0,\ 
\Delta=\Delta^{a}_{a},\ (ab)=\frac{1}{2}(ab+ba).
\end{eqnarray*}

The most interesting part is the skew-symmetric part which
is proportional to vorticity

\begin{eqnarray}
W_{ab}=-R^{2}(1+w)\Theta \omega_{ab},
\end{eqnarray}
\begin{eqnarray*}
w=p/\rho,\ \Theta=3\frac{\dot{R}}{R}.
\end{eqnarray*}

The form of primordial density fluctuations at
decoupling could be fixed as in Ref.\cite{r20}
${\cal D}_{a} \propto A_{a}=\dot{u}_{a}$. Thus we 
only have to follow this component in the evolution 

\begin{eqnarray}
{\cal D}_{a}={\cal D}e_{a},\ e_{a} \parallel A_{a}, 
\end{eqnarray}
\begin{eqnarray*}
\delta = \frac{\delta \rho}{\rho}
\equiv {\cal D},\ e_{a}e^{a}=1,\ e_{a}u^{a}=0.
\end{eqnarray*}

The gauge invariant and covariant formalism assumes
 $\delta_{CMBR}=\delta_{CMBR}(t,x,y,z)$, thus 
within the pure geometric formula we can infer 
some relationships between the cosmological observables 
at large scales
without solving coupled evolution equations \cite{r20a}.
One can evaluate the left- 
and right-hand sides of Eq.(4), giving 
the following relations ($w(CMBR)=\frac{1}{3},\ \vec{X}=R(t)\vec{x}$,
subscript 0 denotes present values):

\begin{eqnarray}
 (\frac{\partial \delta_{CMBR}}{\partial Z})_{0} = 0, \nonumber \\
(\frac{\partial \delta_{CMBR}}{\partial X})_{0}
+\frac{2\omega_{0}}{\sqrt{\Sigma_{0}}}\delta_{CMBR,0}
= 8 R_{0}H_{0}\omega_{0},
\end{eqnarray}
\begin{eqnarray*}
\omega^{2}=\frac{1}{2}\omega_{ab}\omega^{ab},
\ \frac{\delta \rho_{CMBR}}{\rho_{CMBR}}=
4\frac{\delta T_{CMBR}}{T_{CMBR}},
\end{eqnarray*}
\begin{eqnarray*}
\omega_{ab}=\frac{1}{2}h^{c}_{a}h^{d}_{b}
(\nabla_{d}u_{c}-\nabla_{c}u_{d}).
\end{eqnarray*}

In flat geometry, one can define the cosmic scale at present as
the Hubble distance $R_{0}=H_{0}^{-1}$. 
Accounting for the possible relation between expansion and vorticity
as in the Einstein-Cartan cosmology \cite{r3} 
$\omega_{0}\simeq \frac{\sqrt{3}}{2}\Sigma_{0}H_{0}$, 
we obtain

\begin{eqnarray}
H_{0}^{-1}(\frac{\partial \delta_{CMBR}}{\partial X})_{0}
 = \sqrt{3}\Sigma_{0}(4-\frac{\delta_{CMBR,0}}{\sqrt{\Sigma_{0}}}).
\end{eqnarray}

The spatial gradient of the density contrast
 (in the direction perpendicular to the
acceleration vector and the axis of rotation)
  evidently does
 not vanish if the vorticity of the Universe does not vanish, and
vice versa.
By directional continuous spherical wavelets one can
estimate the spatial gradients of the density contrast at large angles and
infer information on the vorticity measured a long
time ago by Birch \cite{r21} ($\omega_{0}={\cal O}(10^{-13}yr^{-1})$)
or calculated from the angular velocity
of galaxies \cite{r22}.

\end{document}